\documentclass[aps, superscriptaddress, prl, twocolumn]{revtex4-1}
\usepackage{ORI_Group_style}

\newcommand{\Ob}{O\bold{e}_1\bold{e}_2\bold{e}_3}
\newcommand{\Ol}{O\uex\uey\uez}

\newcommand{\gr}{\gamma_0}

\newcommand{\Eang}{\hat{\W}}

\newcommand{\eula}{\alpha}
\newcommand{\eulb}{\beta}
\newcommand{\eulc}{\gamma}

\newcommand{\Sup}{\Sop_{\uparrow}}
\newcommand{\Sdw}{\Sop_{\downarrow}}

\newcommand{\Jup}{\Jop_{\uparrow}}
\newcommand{\Jdw}{\Jop_{\downarrow}}

\newcommand{\D}[2]{\hat{\mathcal{D}}_{#1}^{#2}}	
\newcommand{\rop}{\hat{r}}

\newcommand{\jop}{\hat{j}}
\newcommand{\jdop}{\hat{j}^\dag}
\newcommand{\kop}{\hat{k}}
\newcommand{\kdop}{\hat{k}^\dag}
\newcommand{\mop}{\hat{m}}
\newcommand{\mdop}{\hat{m}^\dag}
\newcommand{\bzop}{\hat{b}_z}
\newcommand{\bzdop}{\bzop^\dag}
\newcommand{\brop}{\hat{b}_R}
\newcommand{\brdop}{\hat{b}_R^\dag}
\newcommand{\blop}{\hat{b}_L}
\newcommand{\bldop}{\hat{b}_L^\dag}
\newcommand{\sop}{\hat{s}}
\newcommand{\sdop}{\hat{s}^\dag}

\newcommand{\cop}{\hat{c}}
\newcommand{\cdop}{\cop^\dag}

\newcommand{\Psop}{\hat\Psi}
\newcommand{\Psdop}{\Psop^\dag}
\newcommand{\psop}{\hat{\boldsymbol{\psi}}}
\newcommand{\psdop}{\psop^\dag}
\newcommand{\Phop}{\hat\Phi}

\newcommand{\vphop}{\hat{\boldsymbol{\varphi}}}
\newcommand{\vphdop}{\vphop^\dag}

\newcommand{\wR}{\w_I}
\newcommand{\wD}{\w_D}
\newcommand{\wL}{\w_L}
\newcommand{\wT}{\w_T}
\newcommand{\wZ}{\w_Z}
\newcommand{\wS}{\w_S}

\newcommand{\wmn}{\w_-}
\newcommand{\wpl}{\w_+}

\newcommand{\SJ}{\eta}

\newcommand{\mmu}{\boldsymbol{\mu}}

\begin{document}

\title{Quantum Spin Stabilized Magnetic Levitation}

\author{C. C. Rusconi}
\affiliation{Institute for Quantum Optics and Quantum Information of the
Austrian Academy of Sciences, A-6020 Innsbruck, Austria.}
\affiliation{Institute for Theoretical Physics, University of Innsbruck, A-6020 Innsbruck, Austria.}
\author{V. Pöchhacker}
\affiliation{Institute for Quantum Optics and Quantum Information of the
Austrian Academy of Sciences, A-6020 Innsbruck, Austria.}
\affiliation{Institute for Theoretical Physics, University of Innsbruck, A-6020 Innsbruck, Austria.}
\author{K. Kustura}
\affiliation{Institute for Quantum Optics and Quantum Information of the
Austrian Academy of Sciences, A-6020 Innsbruck, Austria.}
\affiliation{Institute for Theoretical Physics, University of Innsbruck, A-6020 Innsbruck, Austria.}
\author{J. I. Cirac}
\affiliation{Max-Planck-Institut f\"ur Quantenoptik,
Hans-Kopfermann-Str. 1, D-85748, Garching, Germany.}
\author{O. Romero-Isart}
\affiliation{Institute for Quantum Optics and Quantum Information of the
Austrian Academy of Sciences, A-6020 Innsbruck, Austria.}
\affiliation{Institute for Theoretical Physics, University of Innsbruck, A-6020 Innsbruck, Austria.}

\begin{abstract}
We theoretically show that, despite Earnshaw's theorem, a non-rotating single magnetic domain nanoparticle can be stably levitated in an external static magnetic field. The stabilization relies on the quantum spin origin of magnetization, namely the gyromagnetic effect. We predict the existence of two stable phases related to the Einstein--de Haas effect and the Larmor precession.
At a stable point, we derive a quadratic Hamiltonian that describes the quantum fluctuations of the degrees of freedom of the system. We show that in the absence of thermal fluctuations, the quantum state of the nanomagnet at the equilibrium point contains entanglement and squeezing. 

\end{abstract}

\maketitle

According to the Einstein--de Haas and the Barnett effect \citep{EdH,Barnett}, a change in the magnetization of an object is accompanied by a change in its rotational motion. In particular, if the magnetic moment of a magnet is varied by a single Bohr magneton, it must rotate with an angular frequency $\hbar/I$ about the magnetic moment axis to conserve angular momentum. Here $I$ is its moment of inertia about the rotation axis. For a Cobalt sphere of radius $R$, this corresponds to a frequency $\hbar/ I\approx 2\pi \times 10^6 \, \text{Hz}/(R[\text{nm}])^5 $, where $R[\text{nm}]$ is the radius in nanometers. 
This clear manifestation of the quantum spin origin of magnetization, as prescribed by the gyromagnetic relation, is
 hence boosted at the nanoscale~\citep{Quantum_EdH,Chud_EdH1,Chud_EinsteindeHaas}.

In this Letter, we explore the role of the quantum spin origin of magnetization in magnetic levitation.
Earnshaw's theorem~\citep{bassani2006earnshaw}, very relevant in this context, prevents magnetic levitation of a non-rotating ferromagnet in a static magnetic field. The theorem can be circumvented by mechanically spinning the magnet, as neatly demonstrated by the Levitron~\citep{simon1997spin,Dullin,Gov1999214,Berry_Levitron}, which is a magnetic top of a few centimeters. At the single atom level, magnetic trapping with static fields is also possible by exploiting the fast Larmor precession of its quantum spin~\citep{Sukumar1997,Sukumar2006}. In this case, the atom is, from the mechanics point of view, a point particle without rotational degrees of freedom. A magnetic nanoparticle lies in between the Levitron and the atom as both its rotational degrees of freedom and the quantum spin origin of magnetization have to be accounted for. Can a non-rotating magnetic nanoparticle, despite Earnshaw's theorem, be stably levitated with static magnetic fields? 

We show in this Letter that this is the case. In particular, we predict two stabilization mechanisms that crucially rely on the quantum spin origin of the magnetic moment. At low (large) magnetic fields, the Einstein--de Haas effect (the Larmor precession of its magnetic moment) stabilizes levitation. These results are obtained by deriving a quadratic Hamiltonian which describes the linearized dynamics of the degrees of freedom of the magnet (center-of-mass motion, rotation, and magnetization dynamics) around the equilibrium point. We further show that in the absence of thermal fluctuations, the equilibrium state exhibits both quantum entanglement and squeezing of its degrees of freedom. As discussed in the conclusions, these results could be used to bring and control the rich physics of levitated nanomagnets in the quantum regime.

We consider a single magnetic domain nanoparticle (nanomagnet hereafter) in an external static magnetic field $\BB(\rr)$. The nanomagnet is modeled~\citep{Note0} as a spherical rigid body of mass $M$,  radius $R$, and with uniaxial magnetocrystalline anisotropy~\citep{chikazumi}.  The dynamics are described in the body frame $\Ob$,  with $\bold{e}_3$ aligned to the direction of the anisotropy axis. The body frame is obtained from the laboratory frame $\Ol$ by the rotation $R(\W)$, as defined in~\citep{ORI_Cos_PRB}, where $\W=\{\eula,\eulb,\eulc\}$ are the Euler angles. The Hamiltonian of the system in the body frame is given by~\citep{ORI_Cos_PRB}
\be\label{eq:Ham_Body}
\begin{split}
	\Hop =& \frac{\hat{\pp}^2}{2M} + \frac{\hbar^2}{2 I}\pare{\JJop +\SSop}^2-\hbar^2 D\Sop_3^2 +\hbar\gamma_0 \SSop\cdot\BB(\hat \rr,\Eang).\!
\end{split}
\ee
Here  $ \JJop \equiv \LLop + \hat{\mmu}/(\hbar\gr)$ is the total angular momentum (excluding the center-of-mass angular momentum), $\SSop \equiv - \hat{\mmu}/(\hbar\gr) $ the spin angular momentum~\cite{Note3}, $\LLop$ the rigid body angular momentum, $\hat{\mmu}$ the magnetic moment operator, and $\gr>0$ the gyromagnetic ratio \citep{Note1}.
The macrospin approximation is assumed, namely that the total magnetization is constant, which results in $\SSop^2=S(S+1) \id$ with $S \equiv Nf$ for $N$ identical spin $f$ constituents of the nanomagnet \citep{molecularNM}. The first (second) term in \eqnref{eq:Ham_Body} represents the center-of-mass (rotational) kinetic energy of the nanomagnet.  The third term represents the uniaxial magnetocrystalline anisotropy, whose strength is parametrized by  $D\equiv 4\pi R^3k_a/(3\hbar^2 S^2)$ with $k_a$ being the material-dependent anisotropy constant~\cite{molecularNM}. The last term represents the magnetic dipolar interaction, where $\BB(\rr,\W) \equiv R(\W)\BB(\rr)$.  Hereafter, we consider a Ioffe-Pritchard magnetic field, \ie $\BB(\rr) = \uex (B' x -B'' xz /2) - \uey(B' y + B'' yz/2) + \uez [B_0 + B''z^2/2 - B''(x^2+y^2)/4 ]$, where $B_0,B'$ and $B''$  are, respectively, the field bias, gradient and curvature \citep{atomchip}. 
Gravity, which is assumed to be along the $z$-axis, can be safely neglected since throughout the Letter the condition $M g/(\mu B'') \ll (B_0/B'')^{1/2}, (B'/B'')$ is fulfilled, where $g$ is the gravitational acceleration, see \citep{LinStab_NM} for further details. 
The rotational angular momentum of the nanomagnet about the anisotropy axis is a constant of motion, $[\Lop_3,\Hop]=0$~\citep{ORI_Cos_PRB}.
The degrees of freedom of the system, namely the center-of-mass motion, the rotational motion and the magnetization dynamics, are thus described by 12 independent dynamical variables (see~\citep{LinStab_NM} for further details). In the following, we address whether the nanomagnet can stably levitate for a given set of physical parameters of our model: mass density $\rho_M$, magnetization $\rho_\mu$, $k_a$, $R$, $B_0$, $B'$, and $B''$.

\begin{figure}
	\includegraphics[width= 1\columnwidth]{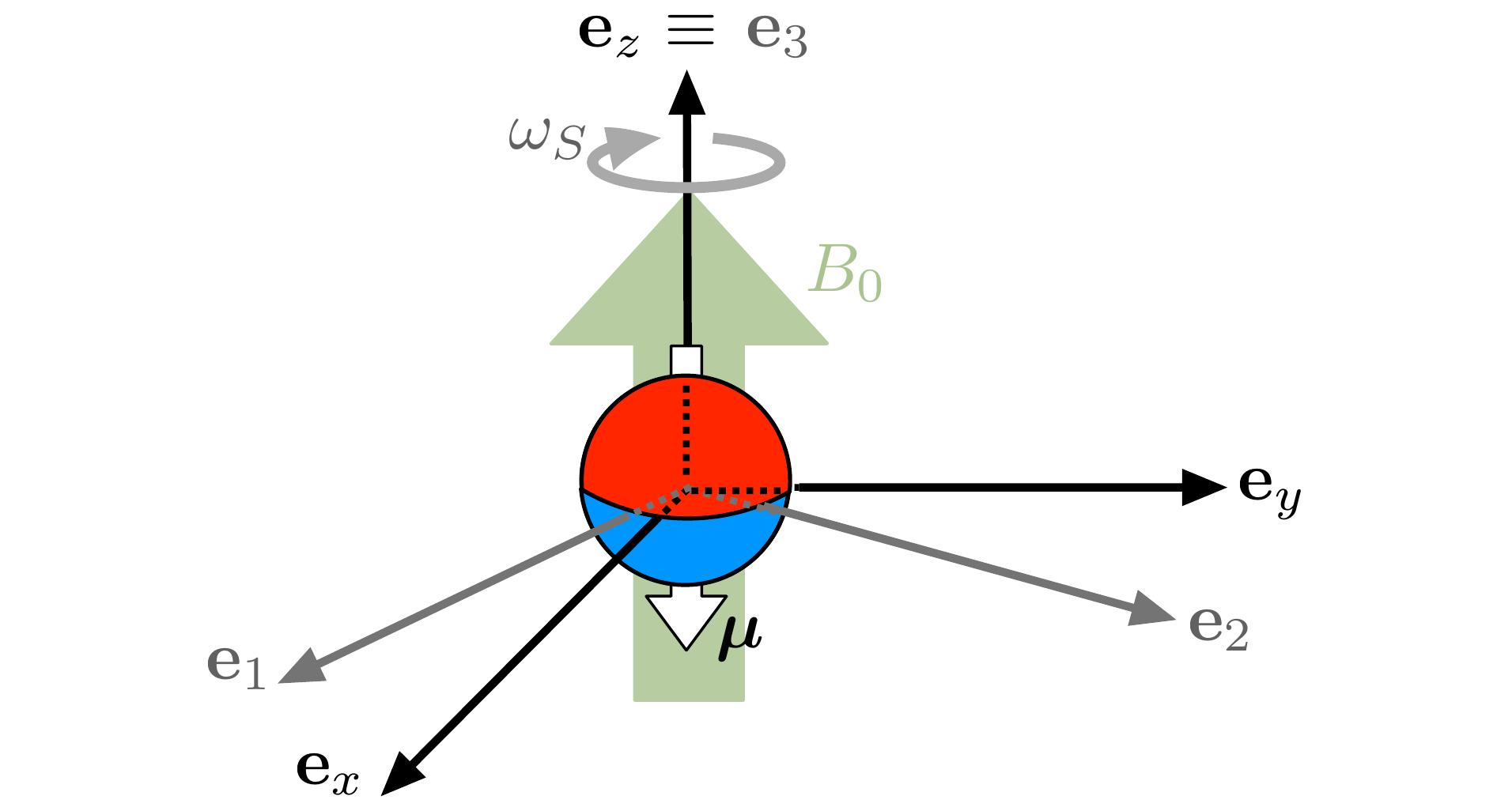}
	\caption{Relative equilibrium for a magnetically levitated nanomagnet.
	The nanomagnet is at the center of the trap and rotates about its magnetization axis at the constant frequency $\wS\equiv -\hbar \avg{\Lop_3}/I$. The magnetic moment lies along the anisotropy axis and is anti-aligned to $\BB(0)=B_0\uez$.}
	\label{Fig:Equilibrium} 
\end{figure}

Let us first obtain the equilibrium configuration of the system. By writing the Heisenberg equations of motion for the nanomagnet in semiclassical approximation~\citep{LinStab_NM}, the following relative equilibrium is found (see~\figref{Fig:Equilibrium}):
(i) The center of mass is fixed at the center of the trap; (ii) The orientation is given by the body frame aligned to the laboratory frame ($\bold{e}_3 \parallel \uez$) and rotating about $\bold{e}_3$ at the frequency $\wS \equiv -\hbar\avg{\Lop_3}/I$; (iii) The magnetic moment lies along the anisotropy axis $\bold{e}_3$ and is anti-aligned to the magnetic field $\BB(0)=B_0\uez$. 
The stability of this relative equilibrium is analyzed in the frame co-rotating with the system~\citep{meiss2007differential,LinStab_NM}. This is obtained via the unitary transformation $\hat{U}=\exp(-\im\wS \Lop_3 t)$ which transforms $\Hop$ into $\Hop+\hbar\wS\Lop_3$.

The linear stability of the system is determined by the dynamics of the fluctuations of the degrees of freedom around the equilibrium. 
The evolution of small fluctuations can be described as a collection of interacting harmonic oscillators through the bosonization procedure presented in~\citep{ORI_Cos_PRB}.
In the following, we provide the mapping between the observables and the bosonic operators and refer to  \citep{ORI_Cos_PRB} for its derivation and further details.
The fluctuations of the center-of-mass motion are described by three independent harmonic oscillators: 
\be
\zop \equiv z_0(\bzop + \bzdop),\hspace{1em}\pop_z \equiv \frac{\im \hbar}{2 z_0}(\bzdop-\bzop),
\ee
for the motion along $\uez$, and 
\be
\rop_+ \equiv  \sqrt{2}r_0 (\brdop+\blop), \hspace{1em}  \pop_+ \equiv \im \frac{\hbar}{\sqrt{2}r_0}(\bldop-\brop),
\ee
for the transverse motion, where $\rop_\pm = \xop\pm\im \yop $ and $\pop_\pm =  \pop_x\pm\im\pop_y $ . Here, 
 $z_0 \equiv \sqrt{\hbar/(2M\wZ)}$ and $r_0 \equiv\sqrt{\hbar/(2M\wT)}$ are the zero-point motions and  $\wZ\equiv \sqrt{\hbar \gr B'' S/M}$ and $\wT\equiv\sqrt{\hbar \gr S (B'^2-B_0B''/2)/MB_0}$ the trap frequencies. We introduced three bosonic modes with $[\bop_i,\bdop_j]=\delta_{ij}$ for $i,j=z,R,L$.
The harmonic oscillators describing the fluctuations of the rotational degrees of freedom are obtained as follows. A Holstein-Primakoff mapping generalized to the case where $\JJop^2$ is not conserved \citep{Marshalek_RMP} leads to 
\bea
\Jup&=&\Jdw^\dag \equiv \Jop_1 -\im \Jop_2 = \sqrt{2J}\kdop, \\
\Jop_3 &=& -J -\sqrt{2J}(\jop+\jdop)/2 -\jdop\jop/2+\kdop\kop.
\eea 
The $k$-bosonic mode ($[\kop,\kdop]=1$) describes the fluctuations of $\Jop_3$ around $\avg{\Jop_3}= -J \equiv -I\wS/\hbar - S$. As discussed below, we assume $J\gg 1$. The $j$-bosonic mode ($[\jop,\jdop]=1$) describes the fluctuations of $\JJop^2$ around $\avg{\JJop^2}= J(J+1)$.
Since the components of $\JJop$ in the laboratory frame commute with $\Jop_3$ and $\Jop_{\downarrow\uparrow}$\citep{Note4}, they need to be bosonized separately as 
\bea
\Jop_+ &=& \Jop_-^\dag \equiv \Jop_x + \im \Jop_y = \sqrt{2J}\mdop, \\
\Jop_z &=& -J -\sqrt{2J}(\jop+\jdop)/2 -\jdop\jop/2+\mdop\mop.
\eea 
The additional $m$-bosonic mode ($[\mop,\mdop]=1$) represents the fluctuations of $\Jop_z$ around $\avg{\Jop_z}=-J$. The bosonization of the Euler angle operators is more involved \citep{ORI_Cos_PRB}.  Since $\Eang$ appears in \eqnref{eq:Ham_Body} only inside trigonometric functions, the Euler angle operators can be written as functions of the 9 Wigner $D$-matrix tensor operators $\D{mk}{1}$ with $m,k=\pm1,0$. The exact bosonization of these operators is given in~\citep{ORI_Cos_PRB} in terms of the $j$-$k$-$m$-bosonic modes.
The fluctuations of the magnetic moment under the macrospin approximation ($\SSop^2=S(S+1) \id$ with $S \gg 1$) can be described by the standard Holstein-Primakoff approximation, namely~\citep{HP_Mapping,Note3}
\bea
\Sup&=&\Sdw^\dag \equiv \Sop_1 - \im \Sop_2 = \sqrt{2S}\sop, \\
\Sop_3&=&S-\sdop\sop.
\eea 
The $s$-bosonic mode describes fluctuations of $\Sop_3$ around $\avg{\Sop_3}=S$. In summary, the fluctuations of the degrees of freedom of the nanomagnet around the equilibrium illustrated in \figref{Fig:Equilibrium} are described by $7$ bosonic modes.

Let us now derive the Hamiltonian describing the dynamics of these bosonic modes. Note that since $\Lop_3=\Jop_3+\Sop_3$ is a constant of motion, the bosonic modes are constrained by $\sqrt{2J} (\jdop+\jop) = \jdop\jop/2-\kdop\kop-\sdop\sop$. Together with the assumption of a slowly rotating nanomagnet, namely $ \vert I\wS/\hbar\vert \ll S$, which implies $J\gg1$, the bosonization procedure transforms $\Hop+\hbar\wS\Lop_3$ into a quadratic Hamiltonian that depends on $6$ bosonic modes: $\bzop,\brop,\blop,\mop,\kop$, and $\sop$. Terms containing the $j$-mode, as well as non-quadratic terms, are of the order $(1/J)^{n/2}$ for $n=1,2\ldots$, and can thus be safely neglected. The quadratic Hamiltonian reads
\be \label{eq:Hq}
\begin{split}
\Hop_q \equiv& \Hop_Z + \Hop_T \\
=&  \frac{\hbar}{2} \pare{\Psdop_Z M_Z  \Psop_Z + \psdop C \psop + \psop C^* \psdop}.
\end{split}
\ee
The first term describes the fluctuations of the center-of-mass motion along $\uez$, where $\Psop_Z = (\bzop,\bzdop)^T$ and $M_Z=\wZ \id_2$.
The second term describes the fluctuations of the other degrees of freedom, where $\psop \equiv (\brdop,\kdop,\blop,\mop,\sop)^T$ and
\be\label{eq:Ham_T}
	C \equiv
	\begin{pmatrix}
	\wmn & g\SJ & -\wpl & -g\SJ & g\\
	g\SJ & \w_k & g\SJ&\wL \SJ^2 & \w_k \SJ^{-1}\\
	-\wpl & g\SJ & \wmn & -g\SJ & g\\
	-g\SJ & \wL \SJ^2 & -g\SJ & -\wL \SJ^2 & \wL\SJ\\
	g & \w_k\SJ^{-1} & g & \wL \SJ & \w_\mu
	\end{pmatrix}.
\ee
Here we defined $\SJ\equiv\sqrt{S/J}$, the Larmor frequency $\wL\equiv\gr B_0$, the anisotropy frequency $\wD\equiv\hbar D S$, the Einstein--de Haas frequency $\wR\equiv \hbar S/I$, the frequencies $\w_k \equiv \wR+\wS -\wL \SJ^2$, $\w_\mu \equiv \wR+2\wD-\wL$, $\w_\pm\equiv(\wT\pm\wZ^2/\wT)/2$, and the coupling strength $g\equiv\wL B'\sigma_T/B_0$. As shown below, the hierarchy between $\wL$, $\wD$, and $\wR$ plays a crucial role in the stability of the nanomagnet.

The linear stability of the equilibrium can be determined by the Heisenberg equations of motion 
\bea
\pa{t} \Psop_Z &=& \frac{1}{\im \hbar}  [\Psop_Z,\Hop_Z] \equiv \im K_Z \Psop_Z=  \im \sigma_z M_Z  \Psop_Z,\label{eq:LinEq_z}\\
\pa{t}\Psop &=& \frac{1}{\im \hbar} [\Psop,\Hop_T] \equiv \im K_T \Psop = \im G M_T \Psop.\label{eq:LinEq_T}
\eea
Here $\sigma_z=\text{diag}(1,-1)$,  $G \equiv \id_5\otimes\sigma_z$, and  $M_\text{T}=M_\text{T}^\dag$ is given by rewriting $\Hop_T$ as $\Hop_T = \hbar\Psdop M_\text{T} \Psop /2$, where  $\Psop \equiv (\brop,\brdop,\blop,\bldop,\mop,\mdop,\kop,\kdop,\sop,\sdop)^T$.
The dynamics of the system are completely characterized by the characteristic polynomials of $\im K_Z$ and $\im K_T$. Linear stability occurs when the roots of the polynomials have zero real part \citep{meiss2007differential}. 
The characteristic polynomial of $\im K_Z$ reads $P_Z(\lambda) = \lambda^2 + \wZ^2$, which has purely imaginary roots for $B''>0$. 
The matrix $\im K_T$ can be block diagonalized in two $5\times 5$ complex conjugate blocks whose characteristic polynomial is given by $P_T(\lambda)=a_0 + a_1 \lambda + a_2\lambda^2 + a_3 \lambda^3 + a_4 \lambda^4 + a_5\lambda^5$, where
\be\label{eq:Poly_Gen_Coeff}
\begin{split}
	a_0 \equiv &\, -2\wD\wR \wL \wT^2, \\
	a_1 \equiv &\, \im\spare{ \wD\wZ^2(\wS + \wR) + \wS \wL \wT^2},\\
	a_2 \equiv &\, -2\wD \wR \wL - \frac{1}{2}\pare{2\wD - \wS}\wZ^2 - \wL\wT^2,\\
	a_3 \equiv & \,\im\spare{-2\wD \pare{\wS + \wR} + \wS \wL + \frac{1}{2}\wZ^2},\\
	a_4 \equiv &\, 2\wD - \wS - \wL,\\
	a_5 \equiv &\, - \im.
\end{split}
\ee
The analysis of $P_Z(\lambda)$ and $P_T(\lambda)$ as functions of the physical parameters of the problem provides the stability diagram of a magnetically levitated nanomagnet, as discussed below. These polynomials can be alternatively obtained either by using classical equations of motion, or via the linearized Heisenberg equations of motion in semiclassical approximation without performing the bosonization, see \citep{LinStab_NM}. Such methods allow to understand the results concerning the stability of a nanomagnet without the need to introduce the bosonization of the quantum mechanical operators. The procedure performed here is richer, since it allows to obtain the quadratic Hamiltonian $\Hop_q$ describing the quantum fluctuations around equilibrium as well as the quantum properties of the states in equilibrium.

\begin{figure}
	\includegraphics[width= 1\columnwidth]{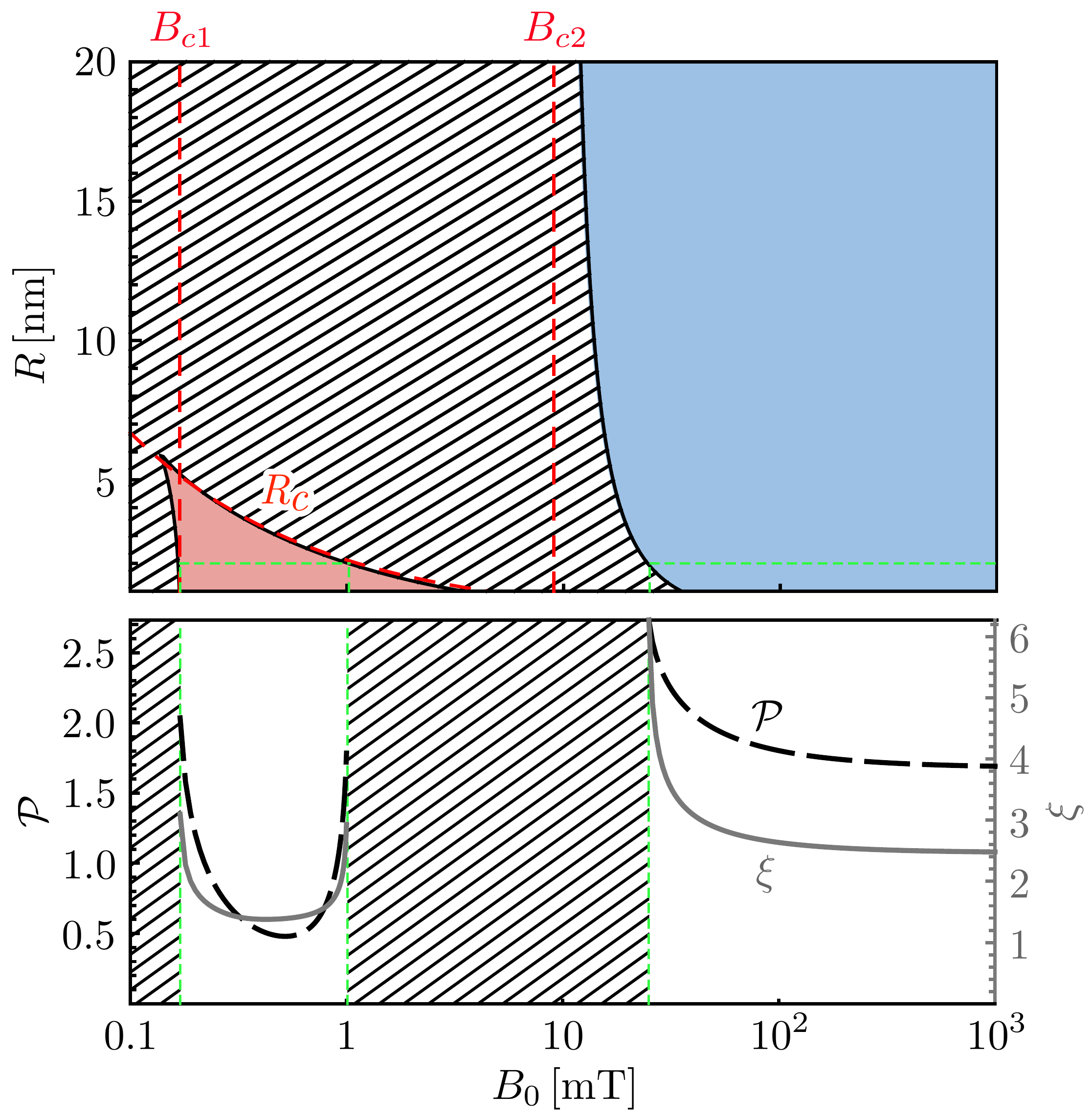}
	\caption{(Top panel) Stability diagram for a non-rotating nanomagnet for $\rho_M = 10^4 \text{Kg}/\text{m}^3$, $\rho_\mu = [\rho_M \mu_B/(50~\text{amu})] \text{J}/(\text{Tm}^3) $($\mu_B$ is the Bohr magneton and amu the atomic mass unit), $k_a = 10^4 \text{J}/\text{m}^3$, $B' = 10^4 \text{T}/\text{m}$, and $B''=10^6 \text{T}/\text{m}^2$. The red-dashed lines represent the approximate borders $B_{c1}=3[\hbar \rho_\mu B'^2/(4\mu_B \gr \rho_M)]^{1/3}$, $R_c \equiv [5\rho_\mu/(8\gr^2 B_0 \rho_M)]^{1/2}$, and $B_{c2}=2k_a \mu_B/(\hbar \gr \rho_\mu)$ of the stable EdH phase (red region) and A phase (blue region).
	(Bottom panel)  Entanglement $\mathcal{P}$ (black dashed line), and squeezing $\xi$ (gray solid line) of the quantum state $\ket{0}$ at the equilibrium as a function of $B_0$ in the stable regions for $R=2\,\text{nm}$ (green dashed lines in the stability diagram).}
	\label{Fig:Plots} 
\end{figure}

Let us now analyze the linear stability of the system in the non-rotating case, see \citep{LinStab_NM} for the $\wS\neq 0$ case.
The stability phase diagram derived using \eqnref{eq:LinEq_z} and \eqnref{eq:LinEq_T}, is shown in~\figref{Fig:Plots}. This is a two-dimensional phase diagram with the $x$-axis given by the bias field $B_0$ and the $y$-axis given by the radius of the nanomagnet $R$. The remaining physical parameters are fixed, see caption of~\figref{Fig:Plots}.
Two stable phases are found.
The Einstein--de Haas (EdH) phase appears at low magnetic fields and for small radius.
In this regime, the dynamics of the magnetization is dominated by the anisotropy interaction ($\wD\gg\wL$).
Due to the small moment-of-inertia-to-magnetic-moment ratio ($\wR\gg\wL$), the angular momentum contribution of the macrospin stabilizes the system through the Einstein--de Haas effect~\citep{EdH}. 
That is, the macrospin is locked along the anisotropy direction due to the conservation of energy~\citep{Chud_EdH1,Chud_RotStatesNM,O'Keeffe_PRB}. Even if rotation is absent, the spin-rotation interplay described by the Einstein--de Haas effect  stabilizes the non-rotating magnet by keeping the macrospin aligned along the anisotropy direction.
The atom (A) phase appears at high magnetic field bias ($\wL\gg\wD$).
In this regime, the nanomagnet behaves like a magnetic atom of mass $M$ and spin $S$~\citep{Sukumar1997,Sukumar2006}: the anisotropy interaction can be neglected and the magnetic moment undergoes a free Larmor precession about the local magnetic field direction. The approximated expressions for the borders of the stable phases, given in the caption of~\figref{Fig:Plots}, can be analytically obtained from the discriminant of the characteristic polynomials~\citep{LinStab_NM}.

The stability diagram shows that a non-rotating nanomagnet can be stably levitated in a static field configuration. This opens up the possibility to cool the thermal fluctuations of the degrees of freedom to the quantum regime. The feasibility and analysis of such an experimental proposal will be addressed elsewhere. Let us now analyze the properties of the quantum state at the equilibrium in the absence of thermal fluctuations.
This state corresponds to the vacuum state $\ket{0}$ of the normal eigenmodes of the quadratic Hamiltonian \eqnref{eq:Hq}: $\bop_Z$ and $\cop_i$ for $i=1,\ldots,5$ ($[\cop_i,\cdop_j]=\delta_{ij}$).  The $c$-bosonic modes $\Phop\equiv(\cop_1,\cdop_1,\ldots,\cop_5,\cdop_5)^T$ are obtained from modes $\Psop$ through a Bogoliubov transformation $\Phop=T^{-1}\Psop$ such that $T^\dag M_T T = \text{diag}(\w_1,\w_1,\ldots,\w_5,\w_5)$ and $T^\dag G T = G$. 
At each stable point of the phase diagram, the transformation $T$ exists~\citep{Maldonado} and can be constructed as follows. One firsts obtains the eigenvalues $\lambda_i$ and eigenvectors $\vv_i$ (for $i=1,\dots,10$) of $K_T$. At a stable point $K_T$ is diagonalizable and  $\lambda_i$ are real and non-degenerate. One then $G$-orthonormalizes the eigenvectors such that, with an appropriate relabeling, they fulfill $\vv_i^\dag G\vv_j=+\delta_{ij}$ for $i,j=1\ldots5$ and $\vv_i^\dag G\vv_j=-\delta_{ij}$ for $ i,j=6\ldots10$~\citep{Note2}.
The Bogoliubov transformation matrix is then given by  $T=(\vv_1\,\uu_1\ldots\vv_5\, \uu_5)$ (the vectors $\vv_i$ and $\uu_i$ are the columns of the matrix), where $\uu_i\equiv(\sigma_x \otimes \mathbb{1}_5)\vv_i^*$ with $\sigma_x$ being the non-diagonal-real-valued Pauli matrix.

One can now analyze the properties of the vacuum state $\ket{0}$, which is a multi-mode Gaussian state, by the $10\times 10$ covariance  matrix~\citep{RSimon}
\be
\Theta_{ij} \equiv \frac{1}{2}\bra{0}(\Psop_i\Psdop_j+\Psdop_j\Psop_i)\ket{0}=\frac{1}{2}\sum_{k=1}^{10} T_{ik}(T^\dag)_{kj}.
\ee
Note that the $\bop_Z$-bosonic mode is not included in $\Theta$ since it is decoupled from all the other modes.
The five $2\times 2$ diagonal blocks $(\Sigma_a)_{ij}\equiv\bra{0}(\vphop_i\vphdop_j+\vphdop_j\vphop_i)\ket{0}/2$ of $\Theta$, with $\vphop\equiv(\aop,\adop)$ and $\aop=\blop,\brop,\mop,\kop,\sop$, correspond to the covariance matrices of the $a$-modes.
The off-diagonal blocks describe the correlations between the modes.
Entanglement in the pure state $\ket{0}$ can be quantified by $\mathcal{P}\equiv 5-\sum_a \mathcal{P}_a$, where $\mathcal{P}_a=[2\sqrt{\text{det}(\Sigma_a)}]^{-1}$~\citep{ParisPurityGss} is the purity of the $a$-bosonic mode. This characterizes the bipartite entanglement between one mode and the remaining four. Single-mode squeezing can be quantified via the squeezing parameter $\xi \equiv1/\sqrt{2\,\text{min}_k({\theta_k})}$, where $\theta_k$ are the eigenvalues of $\Theta$~\citep{RSimon}.
\figref{Fig:Plots} shows $\mathcal{P}$ and $\xi$ as a function of $B_0$ for a given $R$ in the stable phases. Bipartite entanglement and single-mode squeezing are thus present in the $\ket{0}$ state of a magnetically levitated non-rotating nanomagnet.

In conclusion, we showed that the quantum spin origin of the magnetization stabilizes magnetic levitation of a non-rotating nanomagnet, despite the Earnshaw theorem.
Such a quantum-spin-stabilized levitation opens the door to experiments aiming not only at demonstrating the predicted phase diagram, but also at bringing a non-rotating nanomagnet to the quantum regime, whose equilibrium states show non-trivial quantum correlations. There are many directions left for further research, some of which we are currently addressing: The experimental proposal and feasibility analysis to prepare the $\ket{0}$ state by sympathetically cooling the degrees of freedom of a levitated nanomagnet near a flux dependent microwave cavity; The analysis and the potential applications (\eg~spin squeezing for magnetic sensing) of the quantum dynamics generated by placing the nanomagnet at the unstable regions of the phase diagram; Levitating, coupling, and cooling several nanomagnets in the quantum regime to study quantum nanomagnetism in a unique substrate-free environment.
We hope that such results and their potential applications will trigger further theoretical and experimental research to levitate nanomagnets in the quantum regime.

We thank B. Kraus for useful discussions. This work is supported by the European Research Council (ERC-2013-StG 335489 QSuperMag) and the Austrian Federal Ministry of Science, Research, and Economy (BMWFW).

\end{document}